\documentclass[graybox]{svmult}

\usepackage{type1cm}        
\usepackage{makeidx}         
\usepackage{graphicx}        
\usepackage{multicol}        
\usepackage[bottom]{footmisc}

\makeindex             

\begin{document}

\title*{Coupling functions in neuroscience}
\author{Tomislav Stankovski}

\institute{Tomislav Stankovski \at Faculty of Medicine, Ss.\ Cyril and Methodius University in Skopje, Skopje, N.\ Macedonia; Department of Physics, Lancaster University, Lancaster, United Kingdom, \email{t.stankovski@ukim.edu.mk}\\Preprint chapter to the book "\emph{The Physics of Biological Oscillators}"
}
%
%
\maketitle

\abstract{The interactions play one of the central roles in the brain mediating various processes and functions. They are particularly important for the brain as a complex system that has many different functions from the same structural connectivity. When studying such neural interactions the coupling functions are very suitable, as inherently they can reveal the underlaying functional mechanism. This chapter overviews some recent and widely used aspects of coupling functions for studying neural interactions. Coupling functions are discussed in connection to two different levels of brain interactions -- that of neuron interactions and brainwave cross-frequency interactions. Aspects relevant to this from both, theory and methods, are presented. Although the discussion is based on neuroscience, there are strong implications from, and to, other fields as well.       }

\section{Introduction}
\label{sec:1}
Many systems in nature are found to interact, between each other or with the environment. The interaction can cause gradual or sudden changes in their qualitative dynamics, leading to their grouping, self-organizing, clustering, mutual coordinated synchronization, even to some extremes when their very existence is suppressed  \cite{Winfree:67,Haken:83,Kuramoto:84,Pikovsky:01,Strogatz:01b,Stefanovska:07}. An important class of such dynamical systems are oscillators, which also often interact resulting in a quite intricate dynamics.

On the quest to untangle and better understand interactions, one could study several complementary aspects \cite{Friston:11}. One is structural connectivity, where physical actual connection is studied. Often this is not directly observable, or it exist but it is not active and dynamic all the time. Further on, one could study functional connectivity i.e.\ if a functional dependence (like correlation, coherence or mutual information) exist between the observed data. Finally, one could study the causal relations between dynamical models and observe the effective connectivity. In this way, the interactions can be reconstructed in terms of coupling functions  which define the underlaying interaction mechanism.

With their ability to describe the interactions in detail, coupling functions have received a significant attention in the scientific community recently \cite{Stankovski:17b, Stankovski:19}. Three crucial aspects of coupling functions were studied: the theory, methods and applications. Various methods have been designed for reconstruction of coupling functions from data \cite{Kralemann:13b, Kiss:05, Tokuda:07, Levnajic:11, Stankovski:12b, Friston:03}. These have enabled applications in different scientific fields including chemistry \cite{Kiss:07}, climate \cite{Moon:19}, secure communications \cite{Stankovski:14a,Nadzinski:18}, mechanics \cite{Kralemann:08}, social sciences \cite{Ranganathan:14}, and oscillatory interaction in physiology for cardiorespiratory and cardiovascular interactions \cite{Kralemann:13b,Iatsenko:13a,Rosenblum:19b,Lukarski:20,Ticcinelli:17}.

Arguably, the greatest current interest for coupling functions is coming from neuroscience. This is probably because the brain is a highly-connected complex system \cite{Park:13}, with connections on different levels and dimensions, many of them carrying important implications for characteristic neural states and diseases. Coupling functions are particularly appealing here because they can characterize the particular neural mechanisms behind these connections. Recent works have encompassed the theory and inference of a diversity of neural phenomena, levels, physical regions, and physiological conditions \cite{Yeldesbay:19,Onojima:18,Suzuki:18,Eteme:18,Uribarri:19,Stankovski:17c,Su:18,Orio:18,Takembo:18,Stankovski:16,Sanz:18,Bick:20}.

The chapter gives an overview of the topic of coupling function, with particular focus on their use and suitability to neuroscience. This will be explained through observations on two levels of brain connectivity -- the neurons and the brainwaves level. The relationship between the appropriate theory and methods will be also given. On systemic level, the focus will be on neuronal oscillations, thus positioning around and complementing the main topic of the book -- biological oscillators.  The chapter will finish by outlook and some thoughts on the future developments and uses of coupling function in neuroscience. However, before going into greater detail, first the basics of what coupling functions are discussed briefly bellow.

\subsection{Coupling Function Basics}
\label{sec:1.1}

The system setup to be studied is one of an interacting dynamical systems, with the focus of coupled oscillators. Then,
\emph{coupling functions describe the physical rule specifying how the interactions occur and manifest}. Because they are directly connected with the functional dependencies,  coupling functions focus not only on \emph{if} the interactions exist, but more on \emph{how} they appear and develop. For example, when studying phase dynamics of coupled oscillators the magnitude of the phase coupling function affects directly the oscillatory frequency and will describe how the oscillations are being accelerated or decelerated by the influence of the other oscillator. Similarly, if one considers the amplitude dynamics of interacting dynamical systems, the magnitude of coupling function will prescribe how the amplitude is increased or decreased due to the interaction.

First we consider two coupled dynamical systems given in the following general form:
\begin{eqnarray}
\dot{x} &=& f_1(x) + g_1(x,y) \nonumber\\
\dot{y} &=& f_2(y) + g_2(x,y),
\end{eqnarray}
where the functions $f_1(x)$ and $f_2(y)$ describe the inner dynamics, while $g_1(x,y)$ and $g_2(x,y)$ describe the coupling functions in the state space. Then, given that the two dynamical systems are oscillators, and under the assumption that they are weakly nonlinear and weakly coupled, one can apply the phase reduction theory \cite{Kuramoto:84,Nakao:16,Pietras:19}. This yields  simplified approximative systems where the full (at least two dimensional) state space domain is reduced to a one dimensional phase dynamics domain:
\begin{eqnarray}\label{eq:phs}
\dot \phi_1 &=& \omega_1+q_1(\phi_2,\phi_1)\nonumber\\
\dot \phi_2 &=& \omega_2+q_2(\phi_1,\phi_2),
\end{eqnarray}
where $\phi_1,\phi_2$ are the phase variables of the oscillators, $\omega_1,\omega_2$ are their natural frequencies, and $q_1(\phi_2,\phi_1)$ and $q_2(\phi_1,\phi_2)$ \emph{are the coupling functions} in phase dynamics domain. For example, in the Kuramoto model \cite{Kuramoto:84} they were prescribed to be sine functions from the phase differences:
\begin{eqnarray}\label{eq:phs2}
\dot \phi_1 &=&\omega_1+\varepsilon_1\sin(\phi_2-\phi_1)\nonumber\\
\dot \phi_2 &=&\omega_2+\varepsilon_2\sin(\phi_1-\phi_2),
\end{eqnarray}
where $\varepsilon_1,\varepsilon_2$ are the coupling strength parameters. Apart from this example of sinusoidal form, the coupling functions $q_1(\phi_2,\phi_1)$ and $q_2(\phi_1,\phi_2)$ can have very different and more general functional form, including a decomposition on a Fourier series. Given in the phase dynamics like this Eqs.\ \ref{eq:phs}, the coupling functions  $q_1(\phi_2,\phi_1)$ and $q_2(\phi_1,\phi_2)$  are additive to the frequency parameters $\omega_1,\omega_2$, meaning that their higher or lower values will lead to acceleration or deceleration of the affected oscillations, respectively.

Coupling function can be described in terms of its {\it strength} and {\it form}. The coupling strength is a relatively well-studied quantity, and there are many statistical methods which detect measures proportional to it (e.g.\ the mutual-information based measures, transfer entropy and Granger causality). It is the functional form of the coupling function, however, that has provided a new dimension and perspective probing directly the mechanisms of the interactions.  Where, the \emph{mechanism} is defined by the functional form that gives the rule and process through which the input values are translated into output values i.e.\ for the interactions it prescribes how the input influence from one system is translated into the output effect on the affected or the coupled system.

In this way a coupling function can describe the qualitative transitions between distinct states of the systems e.g.\ routes into and out of synchronization, oscillation death or network clustering. Moreover, depending on the known form of the coupling function and the detected quantitative inputs, one can even predict transitions to synchronization.
Decomposition of a coupling function provides a description of the functional contributions from each separate subsystem within the coupling relationship. Hence,   by describing the mechanisms, coupling functions reveal more than just investigating correlations and statistical effects.

\section{Suitability of Coupling Functions for Neuroscience}
\label{sec:2}
The human brain is an intriguing organ, considered to be one of the most complex systems in the universe. The adult human brain is estimated to contain 86$\pm$8 billion neurons, with a roughly equal number (85$\pm$10 billion) of non-neuronal cells \cite{Azevedo:09}. Out of these neurons, 16 billion (19\%) are located in the cerebral cortex, and 69 billion (80\%) are in the cerebellum. One of the main features of the brain is how the neurons are connected, and when and how they are active in order to process information and to produce various functionalities.

In neuroscience, the brain connectivity is classified in three different types of connectivity. That is, the brain connectivity refers to a pattern of  links ("structural, or anatomical, connectivity"), of statistical dependencies ("functional connectivity") or of causal model interactions ("effective connectivity") between distinct units within a nervous system \cite{Horwitz:03,Rubinov:10,Friston:11}.  In terms of graph theory of the brain, the units correspond to nodes, while the connectivity links to edges \cite{He:10}. The connectivity pattern between the units is formed by structural links such as synapses or fiber pathways, or it represents statistical or causal relationships measured as cross-correlations, coherence, information flow or the all-important \emph{coupling function}. In this way, therefore, the brain connectivity is  crucial to understand how neurons and neural networks process information.

\begin{figure}
\sidecaption
\includegraphics[scale=.7]{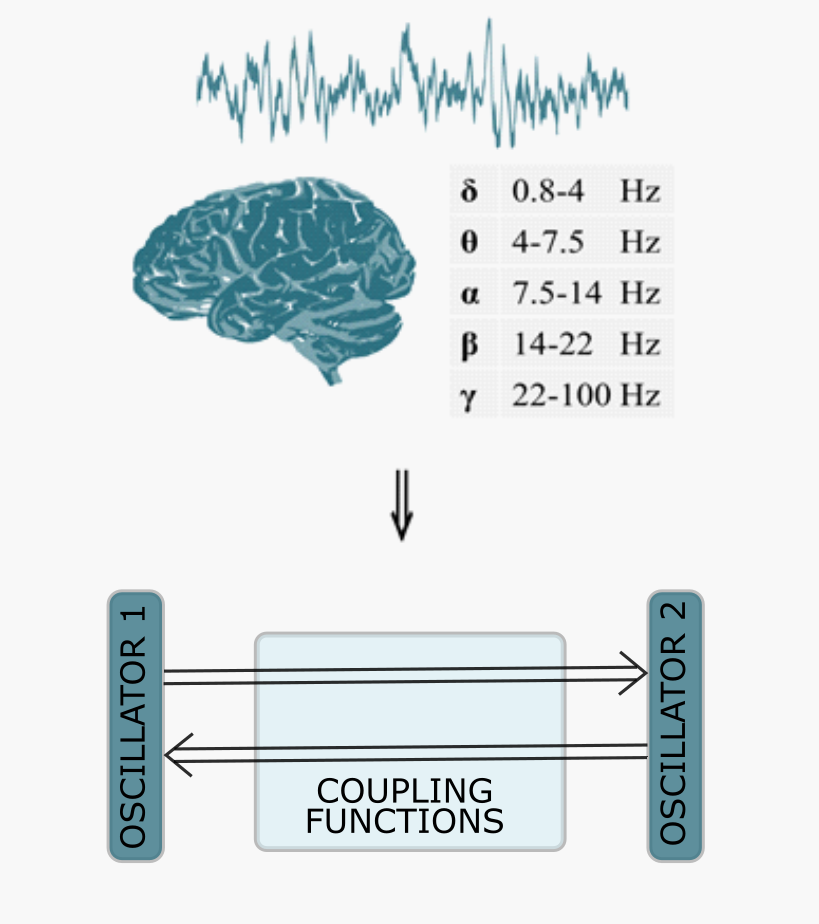}
%
%
\caption{A schematic example of the brain, an electroencephalography (EEG) signal recording as a measure of the neural population electrical activity, and the schematic model of two oscillators and their coupling functions which can be used to model a particular brainwave activity. The five distinct brainwave ($\delta, \theta, \alpha, \beta, \gamma$) frequency intervals are also given on the right of the figure. }
\label{fig:1}       
\end{figure}

The units can correspond to individual neurons, neuronal populations, or anatomically segregated brain regions. Taking aside the anatomically structural brain regions, the other two -- the neurons and their populations -- are of particular interest from a system neuroscience point of view. Moreover, for certain conditions these systems may operate in the oscillatory regime for some time. When having an oscillatory nature their dynamics and connectivity can be modeled as coupled oscillators (see for example Fig.\ \ref{fig:1}). In this constellation, a coupling function with its functional form can be very suitable effective connectivity measure through which much can be learned about the mechanisms and functionality of the brain.

As the two connectivity units, the neurons and the neuronal populations, are of particular interest to the focus of coupling functions and oscillatory dynamics, bellow they will be discussed separately in light of the utility of coupling functions.

\subsection{Coupling Functions on Neuronal Level}
\label{sec:2.1}

The neurons are archetypical cells which act as basic units from which the structure of the brain is realized. Existing in great numbers, they are interconnected in various network configurations giving rise to different functions of the brain. One should note that besides neurons other cell types may also contribute to the brain overall function \cite{Haseloff:05}. As such the brain is a complex system which can perform large number of neural functions from relatively static structure \cite{Park:13}. For comparison, in terms of functions the brain is much more complex than for example the heart, which performs generally only one function -- pumping blood to other parts of the body. Importantly for the brain, the neurons are electrically excitable cells, which are active only in the act of performing certain function.

Based on their function, neurons are typically classified into three types: sensory neurons, motor neurons and interneurons. Number of neuron \emph{models} exist which describe various features, including but not limited to the Hodgkin–Huxley, the Integrate-and-fire, the FitzHugh–Nagumo, the Morris–Lecar and the Izhikevich neuronal model \cite{Gerstner:09,Hodgkin:52,Rocsoreanu:12,Morris:81,Izhikevich:03}.  These models describe the relationship between neuronal membrane electrical currents at the input stage, and membrane voltage at the output stage. Notably, the most extensive experimental description in this category of models was made by Hodgkin–Huxley \cite{Hodgkin:52}, which received the 1963 Nobel Prize in Physiology or Medicine. The mathematical description of the neuronal models is usually represented by a set of ordinary or stochastic differential equations, describing dynamical systems  which under specific conditions exhibit nonlinear \emph{oscillatory} dynamics.

Importantly, the neurons are highly interconnected forming a complex brain network. Their interactions give rise to different neural states and functions. In terms of system interactions, such brain interactions could lead to qualitative transitions like synchronization and clustering, on the whole or part of the brain network. When observing the neuronal models as dynamical systems, the mechanisms of the interactions are defined by the neuronal coupling functions. On this level, coupling functions have been studied extensively, although more in an indirect way through the neuronal phase response curve (PRC) \cite{Ermentrout:96,Achuthan:09}. Namely, coupling function is a convolution between two functions, the phase response curve and the perturbation function \cite{Kuramoto:84} i.e.\ one function of how an oscillator responds to perturbations and the second function defining the perturbations from the second oscillator, respectively. There are generally two types of such response curves, type I with all positive, and type II with positive and negative values. Different types of phase response curves were studied (especially theoretically) forming different types of neuronal models \cite{Brown:04,Ermentrout:06,Oprisan:04}. The  phase
response curves are typically defined for weakly coupled units \cite{Pietras:19,Nakao:16}.

An important feature of the neuronal oscillations are that they are excitable and have non-smooth spike-like trajectories. Such dynamics of the neuronal oscillations are highly nonlinear. For many applications, the neuronal activity is studied completely through the timing of the spike events \cite{Gerstner:02}. In general, such spike-like oscillations act similar as a delta function, hence the phase response curves will have a similar delta function-like form \cite{Ermentrout:12}. This can have direct effect when observing the coupling function which can be a convolution between the a delta-like functions.

In terms of methods for neuronal coupling functions, a number of methods exist for reconstructing the neuronal phase response curves and the associate coupling functions \cite{Galan:05,Suzuki:18}. However, there are many open problems on this task and many applications on different types of signals from interacting neurons are yet to be resolved.

\subsection{Coupling Functions on Brainwave Level}
\label{sec:2.2}

Studying some kind of property of a large number of neurons at once, as a whole or region of the brain, scales up the observation on higher level. In this way the resultant measurement of the brain, or region of the brain, is in a way some kind of mean field, a sum of all the functional activities of the individual neurons in a group, ensemble or network. For example such measurements include the neural EEG, iEEG, NIRS,  MRI, CT and PET, which measure different characteristics like the electrical activity, the hemodynamic activity, the perfusion etc. of the whole brain or on specific spatially localized brain regions.

Arguably, the most used high level observable is the EEG. Electroencephalography (EEG) is a noninvasive electrophysiological monitoring method to record electrical activity of the brain. EEG measures voltage fluctuations resulting from ionic current within the neurons of the brain \cite{Niedermeyer:05}. EEG measures electrical activity over a period of time, usually recorded from multiple electrodes placed on the scalp according to some widely accepted protocols, like the International 10–20 system \cite{Jurcak:07} (internationally recognized protocol to describe and apply the location of scalp electrodes).

At first sight the EEG signal looks random-like and complex (see e.g. Fig.\ \ref{fig:1}), however, a detail spectral analysis reveals that there are number of distinct oscillating intervals -- called \emph{brainwaves}. The most commonly studied brainwaves include the delta $\delta$, theta $\theta$, alpha $\alpha$, beta $\beta$ and gamma $\gamma$ neural oscillation \cite{Buzsaki:04}. The frequancy intervals of these brainwaves are also given in Fig.\ \ref{fig:1}. Apart from these, there are also other brainwaves, including the mu $\mu$, faster gamma1 $\gamma_1$ and gamma2 $\gamma_2$ brainwaves, and other more characteristic oscillations like the sleep spindles, thalamocortical oscillations, subthreshold membrane potential oscillations, cardiac cycle etc. The brainwaves are often linked to specific brain functions and mechanisms, though not all of them are known and they are still very active field of research. The existence and strength of the brainwave oscillations are usually determined by spectral Fourier or Wavelet analysis.

The brainwave oscillations emanate from the dynamics of large-scale cell ensembles which oscillate synchronously within characteristic frequency intervals. The different ensembles communicate with each other to integrate their local information flows into a common brain network. One of the most appropriate ways of describing communication of that kind is through \emph{cross-frequency coupling}, and there has been a large number of such studies in recent years to elucidate the functional activity of the brain underlying e.g., cognition, attention, learning and working memory \cite{Jensen:07,Musizza:07,Jirsa:13,Canolty:10,Voytek:10}. The different types of cross-frequency coupling depend on the dynamical properties of the oscillating systems that are coupled, e.g., phase, amplitude/power and frequency, and different combinations of brainwaves have been investigated, including often the $\delta$-$\alpha$, $\theta$-$\gamma$ and $\alpha$-$\gamma$ cross-frequency coupling relation. These types of investigation are usually based on the statistics of the cross-frequency relationship e.g., in terms of correlation or phase-locking, or on a quantification of the coupling amplitude.

Recently, a new type of measure for brain interactions was introduced called \emph{neural cross-frequency coupling functions} \cite{Stankovski:17c}. This measure is one of the central aspects in this chapter. The neural cross-frequency coupling functions describe interactions which are cross-frequency coupling i.e.\ between brainwaves but now describing not only the coupling existence and strength but also the form of coupling function. This functional form acts as another dimension of the coupling with the ability to describe the mechanisms, or the functional law, of the underlaying coupling connection in question \cite{Stankovski:17b}. In simple words, not only \emph{if}, but also \emph{how} the neural coupling takes place.

\begin{figure}
\sidecaption
\includegraphics[scale=.72]{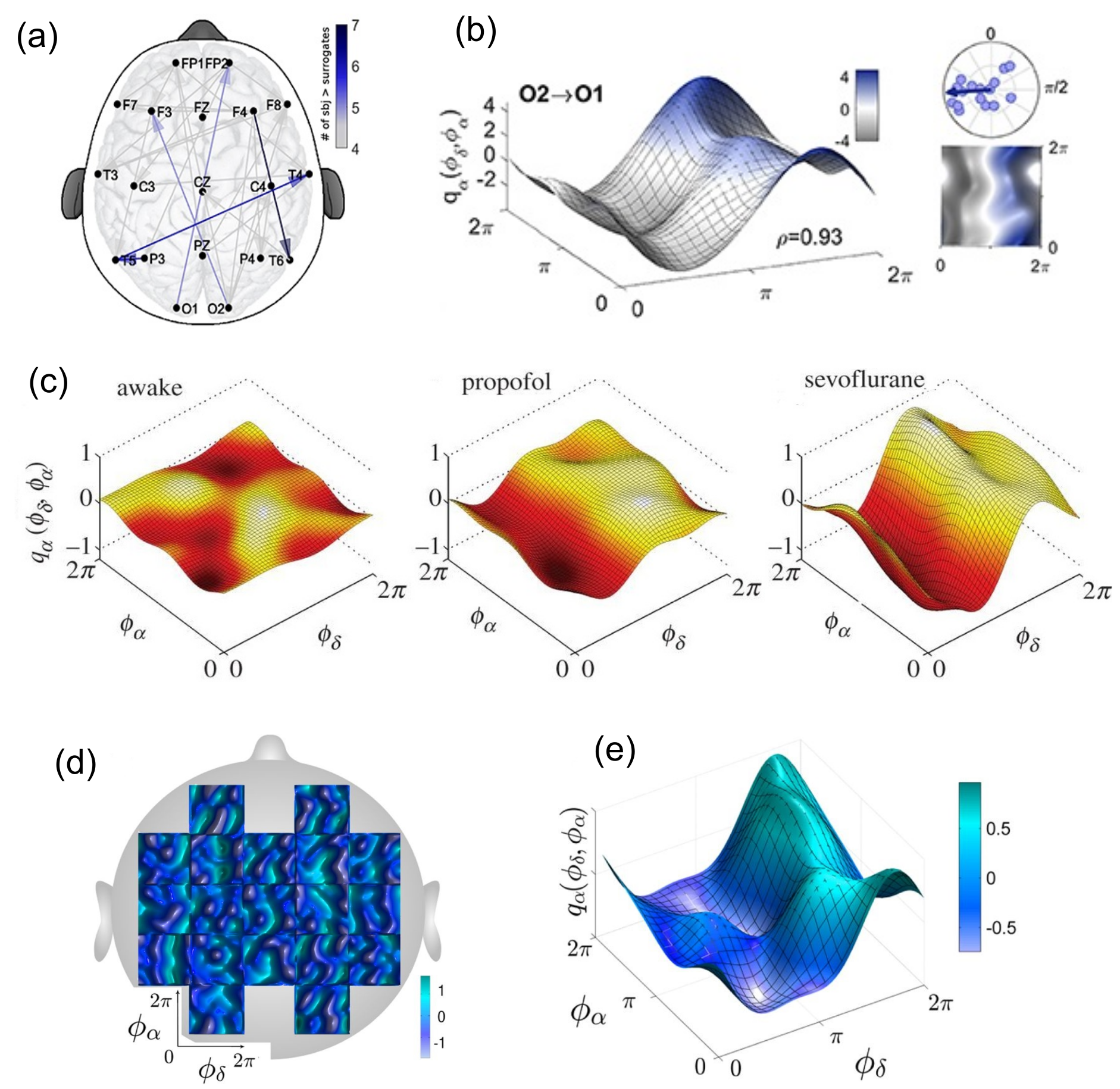}
\caption{Examples of $\delta$-$\alpha$ neural coupling functions. (a) The coupling strength spatial distribution and significance in respect of surrogates. (b) The $\delta$-$\alpha$ phase coupling functions of resting state -- with 3D and 2D plots and a polar index of coupling function similarity.  (c) The effect of anaesthesia on the $\delta$-$\alpha$ coupling functions -- the three group functions are for the awake, anesthetised with propofol and anesthetized with sevoflurane states. (d) and (e) depict the spatial distribution and the average  resting state $\delta$-$\alpha$ coupling function, respectively. (a) and (b) are from \cite{Stankovski:17c}, (c) is from \cite{Stankovski:16}, (d) and (e) are from \cite{Stankovski:15a}.  }
\label{fig:2}       
\end{figure}

When studying brainwave interactions the neural cross-frequency coupling functions are very suitable. Namely, the fact that the brainwaves are described by oscillations can be used to model the interacting dynamics with the coupled phase oscillator model \cite{Kuramoto:84}. In this way one can have a direct 1:1 correspondence between the number of observables and the dimensions of the measured signals -- having a 1D signal and 1D model for the phase dynamics for each system i.e.\ there will be no hidden dimensions. To illustrate the steps of the analysis an example of $\delta$-to-$\alpha$ phase neural coupling function is considered:
\begin{itemize}
  \item First one needs to extract the $\delta$ and $\alpha$ oscillation signals -- this is done with standard filtering of the EEG signals.
  \item After this, one needs to detect the instantaneous phase signals from the oscillations, which can be done by Hilbert transform, and further transforming this with protophase-to-phase transformation \cite{Kralemann:08}.
  \item  Such phases $\phi_{\delta}(t)$ and $\phi_{\alpha}(t)$ are then inputs to a method for dynamical inference which can infer a model of two coupled phase oscillators where the base functions are represented by Fourier series (set of sine and cosine functions of the $\phi_{\delta}(t)$ and $\phi_{\alpha}(t)$ arguments). In our calculations we used the method for dynamical Bayesian inference \cite{Stankovski:12b} and Fourier series as base function up to the second order.
  \item The resulting inferred model explicitly gives the desired neural coupling functions.
  \item After reconstructing the neural coupling functions of interest, one can use them to perform coupling function analysis in order to extract and quantify unique characteristics.
\end{itemize}

The phase coupling functions give the precise mechanism of how one oscillation is accelerated or decelerated as an effect of another oscillation. For example, lets consider the $\delta$-to-$\alpha$ phase neural coupling function. Fig.\ \ref{fig:2} presents such $\delta$-to-$\alpha$ coupling function $q_{\alpha}(\phi_{\delta}(t),\phi_{\alpha}(t))$ from three studies involving resting state and anaesthesia, from single electrode or from spatially distributed electrodes \cite{Stankovski:17c,Stankovski:16,Stankovski:15a}. Fig.\ \ref{fig:2} (a) shows the coupling existence, strength and significance in respect of surrogates, while the  Fig.\ \ref{fig:2} (b) shows the all-important neural coupling function $q_{\alpha}(\phi_{\delta}(t),\phi_{\alpha}(t))$. Observing closely the 3D plot in Fig.\ \ref{fig:2} describes that the $q_{\alpha}(\phi_{\delta}(t),\phi_{\alpha}(t))$ coupling function which is evaluated in the $\phi_{\alpha}(t)$ dynamics changes mostly along the $\phi_{\delta}(t)$ axis, meaning it is a predominantly direct coupling from $\delta$ oscillations. Detailed description of the direct form of coupling function, which is not analytical for non-parametric functional form, are presented elsewhere \cite{Stankovski:15a}. The specific form of the coupling function describes the coupling mechanism that when the $\delta$ oscillations are between 0 and $\pi$ the coupling function is negative and the $\alpha$ oscillations are decelerated, while when the $\delta$ oscillations are between $\pi$ and $2\pi$ the coupling function is positive and the $\alpha$ oscillations are accelerated. The rest of the figures tell similar story -- Fig.\ \ref{fig:2} (c) present three cases of $q_{\alpha}(\phi_{\delta}(t),\phi_{\alpha}(t))$ coupling functions for awake and anaesthetized subjects (with propofol nd sevoflurane anaesthetics, respectively), while Fig.\ \ref{fig:2} (d) and (e) present the $q_{\alpha}(\phi_{\delta}(t),\phi_{\alpha}(t))$ in spatial distribution on the cortex and its average value. The 3D plots present the qualitative description, while for quantitative analysis one can extract two measures -- the coupling strength and the similarity of form of coupling function \cite{Stankovski:17b}.

\section{Theory and Methods for Coupling Functions in Neuroscience}
\label{sec:3}

The theory and methods for studying coupling functions of brain interactions are developed unsymmetrically. Namely, it seems that theoretical studies are more developed for the neuronal level, while the methods are largely developed for studying the large-scale (brainwaves) systems. Of course, this is not a black-and-white division, however the predominance of the two aspects certainly seems to be like this.

The large populations of interacting neurons, in form of ensembles and networks, have been studied extensively in theory. The celebrated Kuramoto model \cite{Kuramoto:75,Kuramoto:84} has been exploited in particular. It is a model of large population of phase oscillators, one which has an exact analytic solution for the synchronization state of the whole ensemble. The coupling functions is a simple sine function of the phase difference. Kuramoto discussed that this coupling function is not very physical, however his interest was in finding an analytically solvable model. The Kuramoto model has been particularly popular in neuroscience with its ability to describe analytically the synchronous states of large populations of neurons \cite{Breakspear:10,Petkoski:19,Acebron:05}. Other two recently introduced approaches, known as the Ott–Antonsen \cite{Ott:08} and Watanabe–Strogatz \cite{Watanabe:93} reductions, provide reduced model equations  that exactly describe the collective dynamics for each subpopulation in the neural oscillator network via few collective variables only. A recent review provides a comprehensive and updated overview on the topic \cite{Bick:20}. The theoretical studies on the large-scale brainwave interactions are often performed through the common framework of two or few coupled oscillatory systems \cite{Pikovsky:01}.

To infer coupling functions from data one needs to employ methods based on dynamical inference. These are class of methods which can reconstruct a model of ordinary or stochastic differential equations from data. The coupling functions are integral part of such models. In this chapter example were shown from the use of specific method based on dynamical Bayesian inference \cite{Stankovski:12b,Smelyanskiy:05a,Stankovski:14d}, however any other method based on dynamical inference (often referred to also as dynamic modelling or dynamic filtering) can also be used \cite{Kralemann:13b, Kiss:05, Tokuda:07, Levnajic:11, Friston:03}. The differences between the results of these methods in terms of the coupling functions are minor and not qualitatively different. Often, there is a need for coupling functions to be inferred from networks of interacting systems, and several methods have been applied in this way \cite{Pikovsky:18,Stankovski:15a,Levnajic:11}. In neuroscience, such methods have been used mainly on two to several brainwave oscillation systems, and it has been argued that the precision and feasibility are exponentially reduced as the number of systems increases and it is recommended not to go beyond $N>10$ \cite{Rings:16}. For this reason and due to the exponentially increasing demand for larger number of systems, there are not many effective methods for inference of coupling functions in low-level large populations of neuronal interactions.

In terms of methodology and analysis, few other aspects are important when analysing coupling functions. One is that once  coupling functions are inferred they give the qualitative mechanisms but for any quantitative evaluations and comparisons (for example in a multisubject neuroscience study) one can conduct coupling function analysis i.e. it can calculate the coupling strength and the similarity of the form of coupling function \cite{Stankovski:17b,Kralemann:13b,Ticcinelli:17}. Also, of paramount importance is to validate if the inferred coupling functions are statistically significant in respect of surrogate time series \cite{Lancaster:18a,Schreiber:00b}. Usually one test the if the coupling strength of coupling functions is significantly higher than the coupling strength from large number of randomized surrogate time series which have similar statistical properties as the original data. Also one should be careful when analysing neural coupling functions as it has been shown that they can be time-varying \cite{Stankovski:17c,Stankovski:17,Hagos:19}, hence this should be taken into account in the analysis.

\section{Conclusions and Discussions}
\label{sec:4}

In summary, this chapter gives an overview of how coupling functions are relevant and useful in neuroscience. They bring an additional dimension -- the form of coupling function -- which revels the mechanism of the neural interactions. This is relevant in neuroscience, as it can describe and be linked to the many different brain functions.

Two largely studied levels of neural interactions were discussed, the low-level individual neurons and the high-level systemic processes like the brainwave oscillations. Of course, these two levels are not excluding but they are closely related, i.e.\ the brainwaves are like a mean-field averages of activities from billions of neurons. In fact studies exist where the brainwave oscillations are modeled as Kuramoto ensembles but the large-scale cross-frequency couplings for the modelling are inferred from data \cite{Schmidt:14,Breakspear:10}. Needless to say,  coupling functions have implications for other levels and depths of the brain other than the two discussed here.

The focus was on phase coupling functions, though the interactions can be in amplitude, or combine phase-amplitude based domains \cite{Jensen:07,Jirsa:13,Canolty:06}. Many modeling methods used in neuroscience actually inferred dynamical systems where coupling functions were an integral part \cite{Jafarian:19,Friston:03}. In such cases coupling functions were implicit, and they were not treated as separate entities, nor were they assesses and analysed separately. These tasks are yet to be developed properly for the amplitude and the phase-amplitude domains.

As an outlook, with all their advantages  one could expect that  coupling functions will continue to play an important role in future neuroscience studies, maybe even to extend their current use. The ever demanding computational power for calculations on large populations of neuron interactions will be more accessible in future, as new improved and faster methods will be developed. The artificial neural networks take on increasing importance recently, with many application across different disciplines and industries \cite{Hassoun:95,Yegnanarayana:09}. The coupling function theory and the different findings in many neuroscience studies could play an important role in establishing improved and more efficient artificial neural networks. Also, the models could be extended and generalized further for easier applications on amplitude and phase-amplitude domains. The theory needs to follow closer the new discoveries from neural coupling functions analysis. The coupling function developments in other fields, especially in physics, could play an important role for neuroscience tasks, and \emph{vice versa}.

%

\bibliographystyle{spphys}

\end{document}